\documentclass{singlecol-new}
\usepackage{natbib,stfloats}
\usepackage{mathrsfs}
\usepackage{graphicx,url}
\usepackage{amsmath,amssymb,amsfonts}
\usepackage{algorithmic}
\usepackage{textcomp}
\usepackage{url}
\usepackage{enumerate}
\usepackage{multirow}
\usepackage{soul, soulutf8}
\usepackage{verbatim}
\usepackage[table,xcdraw]{xcolor}
\usepackage[shortlabels]{enumitem}
\usepackage{scalefnt} % pacote para diminuir fonte das tabelas
\usepackage{comment} % pacote de comentarios
\usepackage{lscape}

\theoremstyle{TH}{

}

\theoremstyle{THrm}{

}

\theoremstyle{THhit}{

}

\makeatletter
\def\theequation{\arabic{equation}}

\makeatother

\begin{document}%
%%%%%%%%%%%%%%%%%

\setcounter{page}{1}

\LRH{M.R. Ferreira et~al.}
%\LRH{Blind Review}

\RRH{Web-based Interactive Narratives to Present Business Processes Models}

\VOL{x}

\ISSUE{x}

\PUBYEAR{202X}

\BottomCatch

\CLline

\subtitle{}

\title{Web-based Interactive Narratives to Present Business Processes Models}

\authorA{Márcio Rocha Ferreira* and \\ Tadeu Moreira de Classe}
\affA{Games to Complex Contexts Research Group (JOCCOM),\\
Graduate Program in Informatics (PPGI),\\ Federal University of the State of Rio de Janeiro (UNIRIO),\\ Rio de Janeiro, RJ, Brazil \\
E-mail: marcio.ferreira@edu.unirio.br\\
E-mail: tadeu.classe@uniriotec.br\\
*Corresponding author}

\authorB{Sean Wolfgand Matsui Siqueira}
\affC{Semantics and Learning Research Group (SaL),\\Graduate Program in Informatics (PPGI),\\ Federal University of the State of Rio de Janeiro (UNIRIO),\\ Rio de Janeiro, RJ, Brazil \\
E-mail: sean@uniriotec.br}

\begin{abstract}
Interactive narratives offer a novel approach to presenting business process models, making them more accessible and collaborative. These narratives create a hyper-textual environment that facilitates knowledge exchange and comprehension for ordinary individuals. However, designing such narratives is complex, as business process modelers must accurately identify and translate the graphic elements of a process model into dynamic narrative elements. This research paper introduces the Scripting Your Process (SYP) method, which provides a systematic approach to designing interactive narratives based on business process models. Following the principles of Design Science Research (DSR), a quasi-experimental study demonstrates and evaluates the SYP method. The results show that the SYP method successfully achieves its objective, contributing to the systematic design of interactive narratives derived from business process models. Consequently, individuals who are not experts in business process management can understand these processes in an engaging and gameful manner.
\end{abstract}

\KEYWORD{Business Process Models; Interactive Narratives; Scripting Your Process; Design Science Research; Business Process Modelling.}

\REF{\textbf{\textcolor{red}{This paper has already been accepted in the International Journal of Business Information Systems (IJBIS) and it is in the journal publication schedule. This paper is the \underline{Author's Original and personal} version.}}}

\maketitle

\section{Introduction}
\label{sec:introducao}

Narrative fictions or interactive fictions are literary genres that allow readers (spectators or even players) to interact and influence the story actively \citep{donikian2004writing}. Interactive narratives can go beyond the purpose of entertainment, being good in other contexts such as game prototyping, teaching and learning and training \citep{montfort2007generating, luo2015review}. We propose interactive narratives to portray business process models.

Organizations use business process management (BPM) concepts and techniques to constantly revise, analyze, and improve their processes \citep{dumas2018fundamentals}. The BPM foresees the business process modeling as a fundamental step to understand how the organization executes its processes, identifying lacks and improvements \citep{engiel2014designing}. In every process modeling language, elements such as symbols and diagrams aim to unify the understanding of the process for managers and people involved in the process context. Nevertheless, it might not be valid for ordinary people (staff, performers, clients, and others) \citep{antunes2020eliciting}. 

When we think about understanding, training, and learning purposes, we believe that interactive narratives may contribute to understanding business processes, presenting features, and simulating their execution \citep{juchova_storyboards_2010}. Using immersion and knowledge sharing from social media and the interaction from hypertext documents, people can interact with processes’ stories, learn processes’ execution, their rules, resources, and characteristics, share perceptions, and think in ideas to improve the organisation \citep{brambilla2012combining}. These narratives represent a business process model in textual language, making these “texts” dynamic, enabling people interaction and, consequently, improving the learning of these processes in a gameful online environment \citep{luo2015review}. 

Although the idea is to make the business process models more uncomplicated and more understandable for people, developing interactive narratives from business process models is not straightforward. The screenwriters responsible for the narrative design may not understand the process model elements and not accurately represent the business process model. They may need some support to perform that task. Therefore, our research question is: \textbf{\textit{How to design web-based interactive narratives from business process models?}} Thus, this research aims to provide a systematic and easy method that guides to less expensive design interactive narratives from business process models with fidelity.

We follow the Design Science Research (DSR) epistemology. The artifact developed to support this research is founded on design requirements and theoretical grounds to solve a problem in a particular context. Thus, the DSR predicts that the artifact evaluation can contribute to scientific and technological knowledge  \citep{hevner2004design}. We proposed the Scripting Your Process (SYP) method as the artifact. We evaluated the artifact using a quasi-experimental approach to analyze whether the SYP method is easy and helpful to screenwriters in narrative design. As a result, we evidenced that the SYP method could solve the problem satisfactorily, being straightforward and helpful in creating coherent interactive narratives based on business process models.

We organized this article as follows: Section \ref{sec:conceitos_fundamentais} presents the research’s conceptual background. Section \ref{sec:trabalhos_relacionados} presents related works. Section \ref{sec:methodology} shows this research in terms of DSR. Section \ref{sec:syp} describes the Scripting Your Process method and a demonstration. Section \ref{sec:evaluation} presents a quasi-experimental study to evaluate whether SYP supports the interactive narrative design from process models entirely and correctly. Finally, Section \ref{sec:conclusion} highlights conclusions and future work.

\section{Backgrounds}
\label{sec:conceitos_fundamentais}

\subsection{Business Processes Modeling}
\label{subsec:bpmn}

Business process modeling is a BPM’s vital step to developing a process model that already exists (\textit{as-is}) or a future process model (\textit{to-be}) to improve it \citep{dumas2018fundamentals}. In general, the modeling uses symbols with semantic meanings to represent process models (process modeling languages – graphic models) \citep{dumas2018fundamentals}. Also, the modeling activity can use speech/writing languages to express process models \citep{santoro_acquiring_2010}. These process models expressed in writing languages show process instances or, even, the view of organizational staff of how the process is performed from their perspectives. Hence, textual process models help process analysts to understand the organizational flow and correctly model the process flow \citep{santoro_acquiring_2010}.

The BPMN language is an international pattern created by \textit{Object Management Group (OMG)} for graphical representations of business process modeling, most used in organizations around the globe \citep{object_management_group_inc_omg_business_2014}. It represents sequence flows, events, decisions, activities, resources, and performers in the business processes. The BPMN language comprises graphic elements with semantic meanings related to the business process contexts \citep{dumas2018fundamentals}. 

On the other hand, textual modeling is beneficial by its simplicity and easy understanding, and it is a natural way that people communicate \citep{de2011let, antunes2020eliciting}. Even though processes languages are made to provide a unique communication form to process managers, their symbols could be complicated for people unfamiliar with them. In this sense, bringing these symbols as narrative elements is an opportunity to innovate in organizations, allowing people to understand and participate in processes improvements \citep{santoro_acquiring_2010, dumas2018fundamentals}. 

Generally, narrative models are processes that represent how an individual (or a group of individuals) performs a process flow. They represent actions, flows, and processes’ conditionals like in graphical models \citep{dumas2018fundamentals}. Commonly, they express some process instances. Additionally, the narrative modeling presents some process’s particularities and characteristics (rules, external data, conditions, roles, and others) that are not explicit in graphical process models. Therefore, people can represent complementary elements easier than in graphic models using narratives models \citep{de2011let}.

\subsection{Interactive Narratives}
\label{subsec:interactive_narratives}

A definition for narratives is the act of telling, reporting, referring to a particular story or happening. Usually, narratives present some essential questions such as a report for or about who, the narrative goal, the style, the environment and others. The narratives arise and evolve simultaneously as the human being’s history with gestures, speeches, texts, multimedia representation, and other ways to tell a story \citep{riedl2013interactive, ryan2017narrative}.

Amid classical literature, cinema, theater, and other passive narratives (stories without interaction), people feel the need to interact and influence the storyline. From that point, interactive narratives or fictions are literary works that allow interaction between the story and spectators \citep{donikian2004writing}.

Interactive narratives are extensions of the classical narratives. They enable people’s direct interaction in the evolution of the story through the spectator’s (readers) choices, actions, and decisions  \citep{donikian2004writing, montfort2005twisty}. For that reason, interactive narratives must be well structured. They should not present flaws because they present only one beginning but could present many ends \citep{luo2015review}.

Some authors consider interactive narratives as a genre derived from cinema and games \citep{donikian2004writing, montfort2005twisty, riedl2013interactive, luo2015review, green2019novella}. For instance, the \textit{Text Games Adventure} emerged in the middle '70s as the first narrative game that enabled interaction \citep{malkewitz2006technologies}. In that context, players usually influence interactive narratives using text commands in a digital environment. In those narratives, players can control characters in a fictional world, send orders to control non-playable characters (NPCs) or manipulate the world's events \citep{montfort2005twisty, riedl2013interactive}.

Nowadays, people can immerse themselves in interactive narratives from social media, games and other communication mechanisms \citep{piredda2015social}. People are not simple spectators in these new forms of telling a story, but they can contribute to the construction of the narrative \citep{mcerlean2018interactive}. Thus, this digital and online environment enabled people to interact with others into narratives, becoming more motivated and collaborative \citep{piredda2015social}.

A narrative (being interactive or not) usually presents some main elements such as narrator, conflict, characters, dramatic actions, plot, time, space, and objects \citep{aarseth_narrative_2012}. Some screenwriters eventually use previous steps such as beat sheets (step outlines) or scripts before creating the complete narrative with all those elements. That practice ensures that the elements and the narrative structure are coherent \citep{aarseth_narrative_2012}. Beat sheets are a way to think about the narrative progression, identifying what will happen in the story step-by-step. Usually, beat sheets present a set of sentences summarizing the main actions. Both beat sheets and scripts encompass the sequence of happenings of a narrative structure \citep{rodriguez2014science}. 

It is fundamental for the excellent result in an interactive narrative project that screenwriters elaborate scripts to analyze the narrative structure and the element coherence. It decreases the risk of the development of a low-quality product.

\section{Related Works}
\label{sec:trabalhos_relacionados}

The related literature presents no method, process, model or framework to support the design of interactive narratives based on business process models. However, we identified some scientific works that use narrative resources to identify business process elements and support modeling the business process \citep{goncalves_collaborative_2010}\citep{de_a_r_goncalves_case_2010}. 

We can exemplify the ``Story Mining'' approach \citep{goncalves_collaborative_2010, de_a_r_goncalves_case_2010} that uses Group Storytelling techniques, allowing people to report business process stories in a collaborative form. That approach also uses natural language processing to extract BPMN elements automatically from texts \citep{goncalves_collaborative_2010}. Using Story Mining, it was possible to map activities from the original process model that traditional methods of information gathering (interviews and others) could not \citep{de_a_r_goncalves_case_2010}. 

Some researchers developed tools to support group storytelling in the business processes modeling, such as the ``ProcessTeller'' \citep{goncalves_collaborative_2010} and the ``TellStory'' \citep{de_a_r_goncalves_case_2010}. Those tools use narrative techniques on business processes texts to make the model activity more precise \citep{goncalves_collaborative_2010}.

\citet{juchova_storyboards_2010} proposed using storyboards(sequence of pictures and multimedia that represents a timeline) as an alternative method for processes models. We understand storyboards as a type of narrative. When we use it, we can describe stories, and in the context of process models, they use the approach to tell the process story. It was a unique work that did not use narrative techniques in the process gathering but for the process presentation.

Concerning alternative methods to represent business process models, there are proposals such as: using games to tell how a process model is \citep{classeEtal2021}; Graphic process model with a detailed explanation \citep{engiel2014designing}; using declarative languages to express process models \citep{giacomo2015declarative} and others. Although these approaches are inspirations to this research work, no one approached the creation of interactive narratives from business process models. The related works use narrative techniques to identify and create business process models, and only one uses an alternative method to present them (ex.: storyboards).

The difference of this research from the other is, precisely, to look into the perspective of extracting elements of a business process model to develop an interactive narrative. No related work researched how to support screenwriters (ordinary people, not process analysts) to emerge the interactive narrative from a process model. Additionally, different from others, we present an alternative form to represent business process models. It is an innovative possibility in business process understanding and training.

\section{Research Methodology}
\label{sec:methodology}

In this research, we used Design Science Research, an epistemological approach to study, research, and explore the artificial (artifacts) to solve problems in a particular context that brings scientific and technological contributions through the artifact evaluation \citep{hevner2004design}. 

\begin{figure}[ht!]
\includegraphics[width=1.0\textwidth]{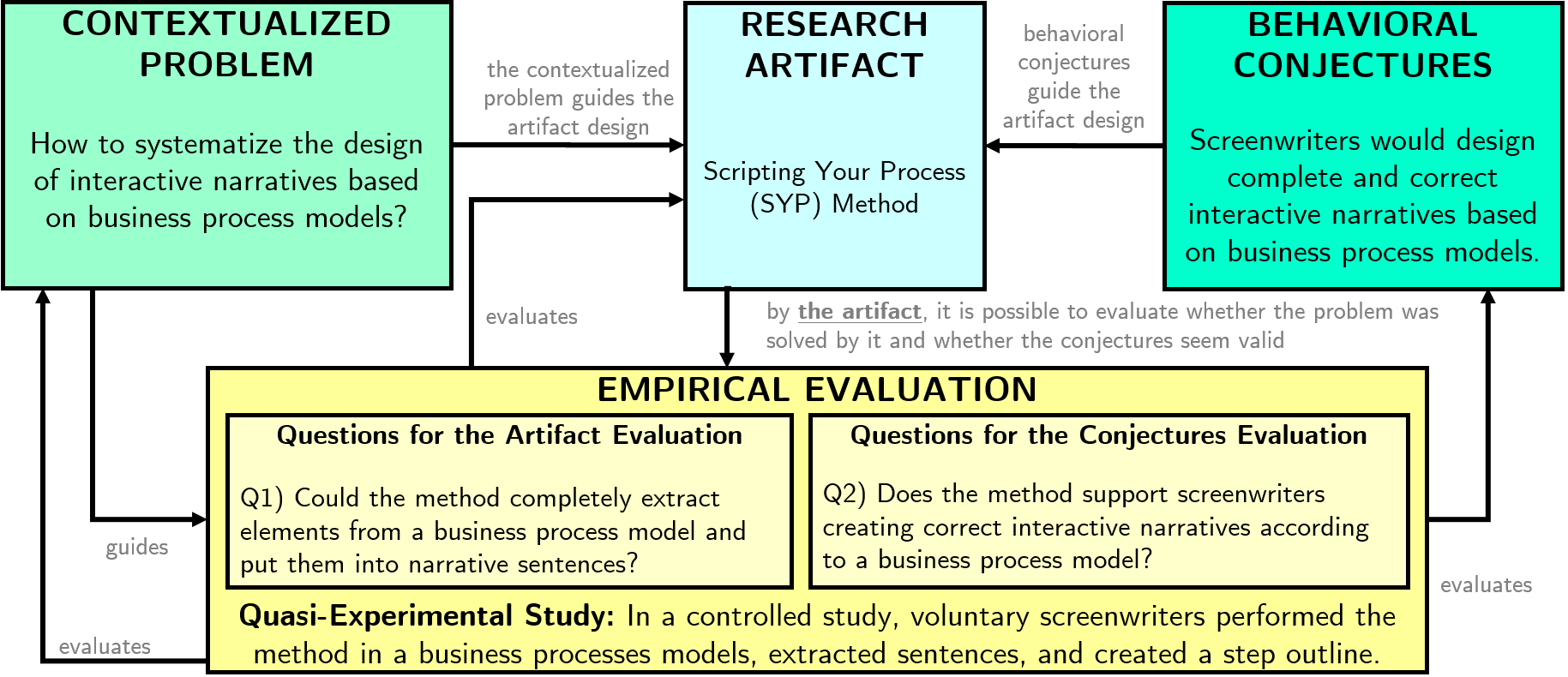}
\caption{Representation of this research, based on central elements of the DSR.}
\label{fig:dsr}
\end{figure}

In DSR, a possible starting point is the problem definition associated with a particular context and audience \citep{pimentel2020design}. The \textbf{contextualized problem} determines the need for an \textbf{artifact} and its requirements. The artifact is designed to solve a problem based on conjectures. The \textbf{behavioral conjectures} assumed affirmations under the people’s behavior (learning, work, communication, and others), related to theories and concepts. An \textbf{empirical evaluation} makes it possible to analyze whether the artifact solved the problem and the conjectures seem valid. Based on that, in Fig. \ref{fig:dsr}, we present this research design in DSR.

The context of this research focuses on ways to make business process models simpler and easier to understand by ordinary people. Thus, we think that the nature of hypermedia, the connected knowledge and the collaboration among people could contribute to creating an interactive narrative that makes the process model simple for everyone. In this case, we can take the interactively from interactive narratives, decreasing the formality of the business process model and contributing to social knowledge.

However, creating interactive narratives is a non-trivial task. Capturing the essence of a business process model and translating it to an interactive narrative demands screenwriters the``know-how'' about process modeling. However, screenwriters are not process analysts, and as ordinary people, they do not need to know about process modeling. Hence, the main problem is \textbf{designing web-based interactive narratives from business process models} and supporting the screenwriters in this task.

State of the art shows storytelling and storyboard strategies to elucidate business process models from textual contents. In this sense, we can use the idea of “extracting elements” from processes’ texts, think in a reverse view (extracting elements from BPMN), and develop an artifact to solve the research problem. Therefore we present the  \textbf{Scripting Your Process (SYP)}, with well-defined steps, easy to follow, that help screenwriters to create interactive narratives that are coherent to a business process model. 

We applied the empirical evaluation in a quasi-experimental study with screenwriters to analyze the questions: \textbf{\textit{Q1) Could the method completely extract elements from a business process model and put them into narrative sentences? Q2) Does the method support screenwriters creating correct interactive narratives according to a business process model?}} We selected this kind of study because, according to Campbell and Stanley\citep{campbell2015experimental}, it is part of a class of studies of empirical nature, less controlled than classical experiments, and without the need to randomize the selection of the participants. Thereby, we consider this kind of evaluation fits our study purpose.

\section{Scripting Your Process (SYP)}
\label{sec:syp}

\begin{figure}[ht!]
\includegraphics[width=1.0\textwidth]{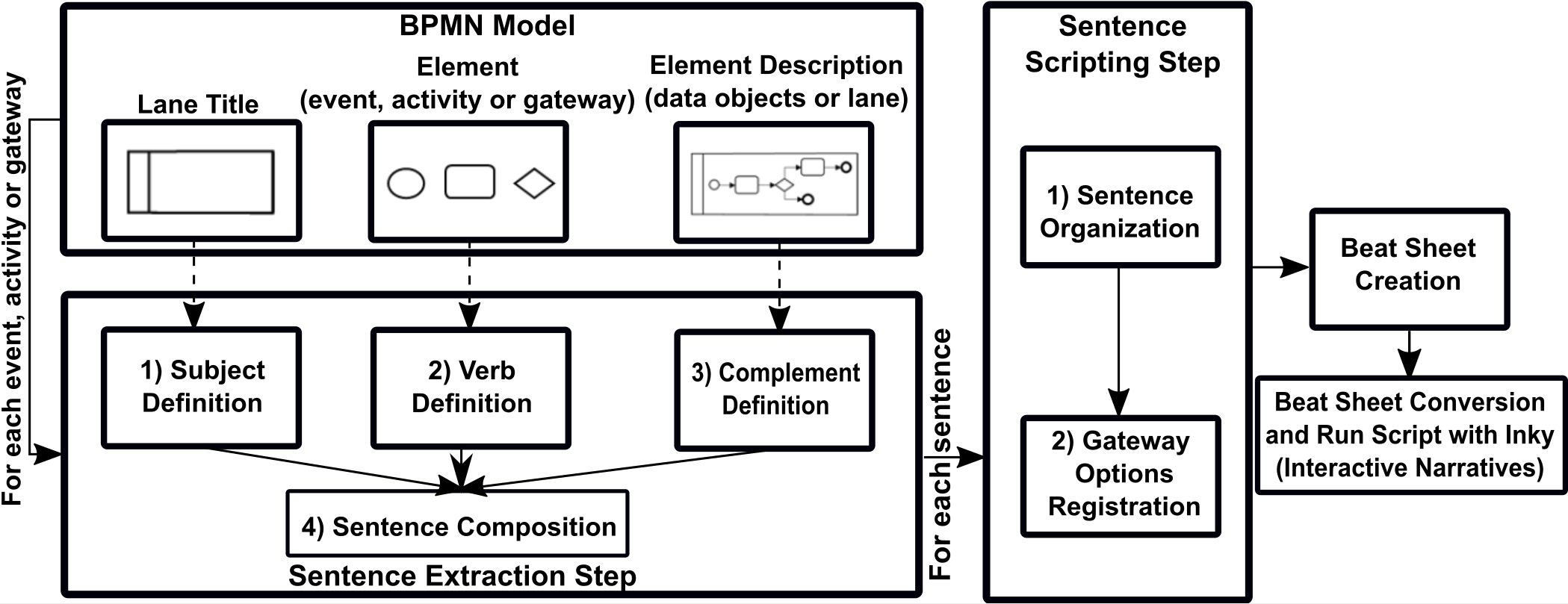}
\caption{Scripting Your Process' steps.}
\label{fig:syp}
\end{figure}

The Scripting Your Process method uses an existing BPMN business process model to generate an interactive narrative. One of its main requirements is that screenwriters associate BPMN elements with narrative elements and organize them into beat sheets. We created the SYP method for screenwriters with no experience with BPMN elements. Focusing on this requirement and thinking of the systematization, the SYP presents two main steps (Fig. \ref{fig:syp}): \textbf{i) sentence extraction} and \textbf{ii) sentences scripting}.

The \textbf{sentence extraction} step involves the creation of sentences in the form: \textbf{\textit{subject + verb + complements}}. This format considers the main elements in a sentence. In the SYP method, the existence of a BPMN flow element (\textbf{event, action or gateway}) determines a sentence because they are associated with happenings, actions, and decisions, respectively, in a business process model  \citep{dumas2018fundamentals}. 

When we analyze \textbf{activities}, the \textbf{subject} of the sentence always will be the \textit{lane} (it is \textbf{who} performs the task -- person, department, system, or other) that contains the activity. In the case of elements of \textit{start and end events}, the SYP defines that the subject of the sentence is the own process (process' description), does not a lane. In both cases, we use \textbf{simple subjects} in the sentence. In elements such as \textit{gateway and intermediate events}, the SYP uses \textbf{undefined subjects} of the sentence because they denote happenings and rules in the process flow. A lane does not (nobody) perform those elements. This approach was necessary to keep the BPMN semantics. In BPMN, those elements have no executors defined by lanes. They are triggered by happening in the process flow.

\textbf{Verbs} are actions, which can vary according to the sentence's subject. In simple subjects (\textit{start/end events and activities}), the verb can match the subject (ex.: the client \textbf{needs} to buy a book). In undefined subjects (\textit{intermediate events and gateways}), the SYP indicates to use verbs in the third person(ex.: \textbf{It is decided} among X or Y). The screenwriters choose the verbs that better match the subject of the sentence.

Lastly, the content of the BPMN elements (\textit{texts or descriptions}) represents the \textbf{complements} of the sentence. The complements can present one or more terms. Usually, an additional term happens when some \textit{resource element} (system, database, document, etc.) connects to an \textit{activity}. In this case, the screenwriter needs to choose a verb to connect resources to other complements of the sentence (ex.: the client \textbf{needs} to buy a book \underline{\textbf{using}} the credit card).

The result of the sentence extraction step is a list of sentences. The total quantity of extracted sentences must be precisely the same as BPMN flow elements (\textit{gateway, activity or event}) in the list. We can verify if the screenwriters could extract all sentences in the BPMN model based on this quantity of sentences.

The \textbf{sentence scripting} step aims to organize the list of sentences in a step outline, considering the sequence of happenings, actions, and business process decisions. In this step, the screenwriters have the opportunity to refine the extracted sentences. They can put the verbs in the correct tense, choose the suitable noun and, make the grammatical structure better.

Also, the SYP method predicts the automatic conversion of the beat sheet to the Ink language\footnote{\url{https://www.inklestudios.com/ink/}} syntax. Ink is a scripting language to create interactive and game narratives. The software Inky\footnote{https://github.com/inkle/inky} is the Ink source code editor and executor, that creates web-based interactive narratives. The execution enables the people involved in the narrative design to check whether the narrative is according to the business process model or is necessary to adjust. The execution also allows people to play the interactive narrative, and with it, they can learn and understand the business process in a ludic form.

\subsection{Demonstration}
\label{subsec:demonstration}

\begin{figure}[ht!]
\includegraphics[width=1.0\textwidth]{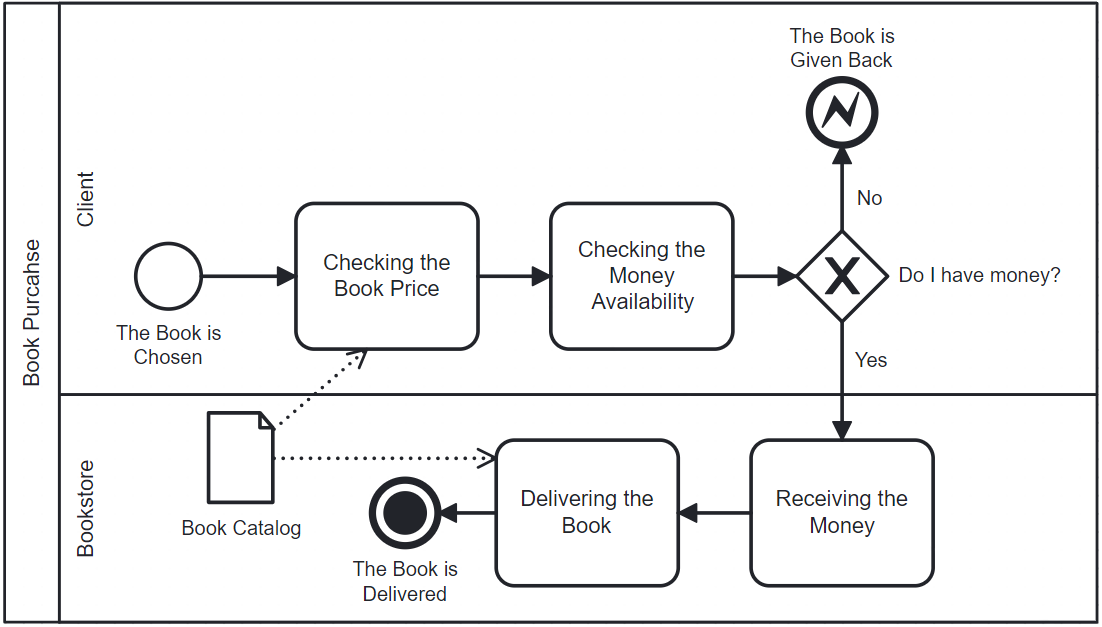}
\caption{Scripting Your Process example of execution}
\label{fig:syp_example}
\end{figure}

Fig. \ref{fig:syp_example}(A) presents an ordinary business process model in the BPMN format. The file BPMN (``.bpmn'') is the primal source for starting the SYP method. In this demonstration, we used a simple business process model that represents the bookstore's buying process. Fig. \ref{fig:syp_example}(B) shows the \textbf{sentence extraction} step happening. Each flow element (\textit{event, activity, and gateway}) originates narrative sentences. Here is possible to see it, eight flow elements originate eight narrative sentences, as predicted in the SYP method. In the following, we will exemplify the possible preliminary extractions from the model presented in Fig. \ref{fig:syp_example}:

\begin{itemize}
    \item \textbf{Start/end events extraction}: \textit{Book Purchase} (process' name -- simple subject) \underline{starts} (verb) when ``The Book is Chosen'' (event description -- complement);
    
    \item \textbf{Activity with resource extraction}: \textit{Client} (lane -- simple subject) \underline{needs} (verb) ``Checking the Book Price'' (activity description -- complement) \underline{using} (verb) ``Book Catalog'' (resource description -- complement);
    
    \item \textbf{Gateway extraction}: \underline{It is decided} (passive voice -- undefined subject) among ``I have money'' (gate option -- complement) OR ``I have no money''(gate option -- complement);
\end{itemize}

The \textbf{sentence scripting} step happens after extracting all sentences. According to the process flow, screenwriters should put the sentences in order based on the list of sentences. Also, they can correct verbs, nouns, and grammatical errors in each sentence. Table \ref{tab:syp_sentences} presents the result of this step. The column ``\#'' is the sentence identification; the next column, ``Sentences'', shows the extracted sentence and improved by the screenwriter; the column ``BPMN element'' is the element that originates the sentence, and the column ``Next'' indicates the following sentence according to the process flow.

\begin{table}[ht!]
\centering
\caption{Sentence scripting from the business process model.}
\label{tab:syp_sentences}
%\resizebox{0.7\textwidth}{!}{
\small
\begin{tabular}{|c|l|c|c|}
\hline
\textbf{\textbf{\#}} & \textbf{\textbf{Sentences}}                                               & \textbf{\textbf{BPMN Element}} & \textbf{\textbf{Next}} \\ \hline
1          & The \textbf{Book Purchase} \underline{starts} when ``The Book is Chosen''                & Start Event           & 2             \\ \hline
2          & The \textbf{Client} \underline{needs} to ``Check the Book Price'' using the ``Book Catalog'' & Activity              & 3             \\ \hline
3          & The \textbf{Client} \underline{needs} to ``Check Its Money Availability''                & Activity              & 4             \\ \hline
4          & \underline{It is Verified} If ``I have money'' OR ``I have no money''               & Gateway               & 6 - 5         \\ \hline
5          & The \textbf{Book Purchase} \underline{ends} when ``The Book is Given Back''              & End Event             & -             \\ \hline
6          & The \textbf{Bookstore} \underline{needs} to ``Receive the Money''                        & Activity              & 7             \\ \hline
7          & The \textbf{Bookstore} \underline{must} ``Deliver the Book'' using the ``Book Catalog''          & Activity              & 8             \\ \hline
8          & The \textbf{Book Purchase} \underline{ends} when ``The Book is Delivered ''              & End Event             & -             \\ \hline
\end{tabular}
%\begin{tablenotes}
%  \tiny
%  \centering
%  \item Note 1: Bold are the simple subjects in the sentences.
%  \item Note 2: Underline are the verbs.
%  \item Note 3: Words in quotations are complements.
%  \item Note 4: Words in standard form are minor adjustments made by screenwriters.
%\end{tablenotes}
%}
\end{table}

\begin{figure}[ht!]
\includegraphics[width=1.0\textwidth]{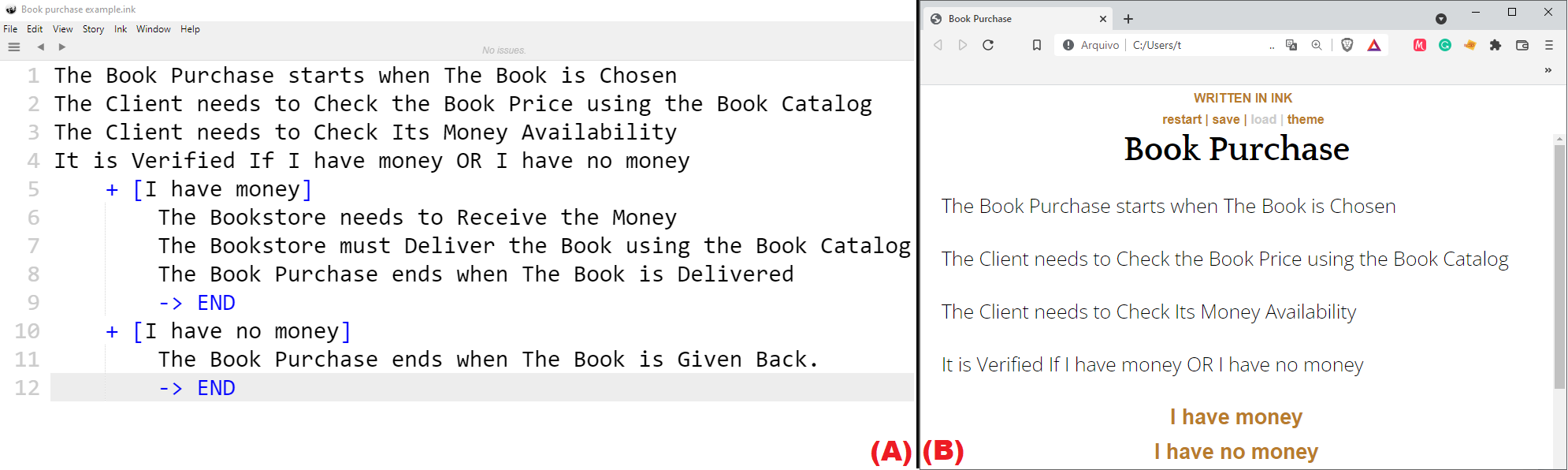}
\caption{Result of converting the list of sentences using the Ink script.}
\label{fig:ink}
\end{figure}

Finally, screenwriters can generate the interactive narrative after the sentence scripting step. Fig. \ref{fig:ink} shows a possible result of converting the list of sentences using the Ink script. On the left side (Fig. \ref{fig:ink}(A)), we presented the source code of the interactive narrative. While on the right (Fig. \ref{fig:ink}(B)), we played the interactive narrative in a web browser. It presents the interactivity elements such as save, reload, restart buttons and hyperlink to narrative decisions. Thus, using the SYP steps, we demonstrated that creating an interactive narrative from a business process model is possible.

\section{Evaluating the Method}
\label{sec:evaluation}

We followed a quasi-experimental study for the empirical evaluation of the artifact. We aimed to analyze whether the SYP method supported screenwriters in the design of interactive narratives based on business process models. According to the DSR, we followed a quasi-experimental study for the empirical evaluation of the artifact. As in classical experiments, quasi-experimental studies follow methodological steps  \citep{campbell2015experimental, kim2016quasi}. In this research, the assessment followed the phases proposed by Campbell and Stanley \citep{campbell2015experimental}: i) definition; ii) planning; iii) execution; iv) data analysis and interpretation; v) discussion of the results.

\subsection{Study Definition}
\label{subsec:definicao_estudo}

We used the GQM (Goal-Question-Metric) paradigm to describe the study definition \citep{basili1992software}. The \textbf{Goal} is: \textbf{Analyze} the SYP method; \textbf{with the purpose of} evaluation; \textbf{regarding} its (1) completeness and (2) correctness; \textbf{from the perspective of} screenwriters; \textbf{in the context of} business process model-based interactive narratives.

According to Fig. \ref{fig:dsr}, the \textbf{Questions} for the empirical evaluation are: \textit{\textbf{Q1) Could the method completely extract elements from a business process model and put them into narrative sentences? Q2) Does the method support screenwriters creating correct interactive narratives according to a business process model?}} 

For each of these questions, we have \textbf{Metrics}. For Q1, we want to \textbf{evaluate if screenwriters could identify and extract all sentences from the business process model}. Thus, the metric used was the number of extracted sentences (QtdExt) by the number of expected sentences (QtdExp): \textbf{$MQ1=QtdExt/QtdExp$}. For Q2, we want to \textbf{observe if the screenwriters extract correct sentences according to the business process model}. So, the metric used was the amount of correct extracted sentences (QtdCorr) by the number of expected sentences (QtdExp): \textbf{$MQ2=QtdCorr/QtdExp$}. We considered the ratio scale for our metrics because it allows us to use descriptive and inferential statistics in our quantitative evaluations. We defined hypotheses for the two research questions to help us interpret the data using inferential statistics (Table \ref{tab:hypothesis}).

\begin{table}[ht!]
\centering
\caption{Hypothesis based on each research question}
\label{tab:hypothesis}
\footnotesize
%\resizebox{\textwidth}{!}{
\small
\begin{tabular}{p{1.5cm}|p{4.9cm}|p{4.9cm}|}
\cline{2-3}
 &
  \textbf{Q1} &
  \textbf{Q2} \\ \hline
\multicolumn{1}{|p{1.2cm}|}{\textbf{Alternative hypothesis}} &
  The SYP method \textbf{CAN COMPLETELY} extracts elements from a BPMN model and puts them into narrative sentences. &
  Using the SYP method process modelers \textbf{CAN CORRECTLY} create interactive narratives from a business process model. \\ \hline
\multicolumn{1}{|p{1.2cm}|}{\textbf{Null hypothesis}} &
  The SYP method \textbf{CAN NOT COMPLETELY} extracts elements from a BPMN model and puts them into narrative sentences. &
  Using the SYP method process modelers \textbf{CAN NOT CORRECTLY} create interactive narratives from a business process model. \\ \hline
\end{tabular}
%}
\end{table}

We believe that mixed studies (qualitative and quantitative) can better understand the results. In this study, beyond the quantitative metrics, we predicted the collection of qualitative data about participants’ perceptions related to the use of the SYP method by a survey. In this sense, the participants’ views, perceptions, and feelings by their qualitative answers complement the quantitative analysis.

\subsection{Study Planning}
\label{subsec:planejamento_estudo}

The planning of the study details the context and limitations of the study. The planning enables a better understanding of the study, allowing its replications and the comparison of its results by scientific peers. For this reason, it is necessary to explain: who the study participants are, what the steps are, what the instrumentation to data collection is, how to treat the data, and the threats of validity.

In this evaluation, we selected the \textbf{participants} by convenience. In total, \textbf{19 participants} attended this research. They were game design students with experience designing narratives for games with a purpose. We invited them to participate as volunteers without offering any gain. For that reason, we considered a good audience for testing the SYP method.

We designed the evaluation in three \textbf{steps}: S1) \textit{Training (30 minutes)}: a training about the steps, features, and execution of the SYP method; S2) \textit{Execution (50 minutes)}: the participants should apply the SYP method in a business process model, extract its sentences and create a step outline that should run in the Inkle; and S3) \textit{Survey (10 minutes)\footnote{\url{https://bit.ly/2UnnOEr}}}: the participants should answer a survey with their perceptions about the method. We designed the study to last 90 minutes maximum.

We defined the \textbf{data instrumentation} based on each study step. For step one (S1), we prepared an SYP guide to support the participants in the following steps. In the second one (S2), we delivered a spreadsheet for sentence extraction and organized them into a step outline. The spreadsheet implements functionalities to export the extracted sentences to the Ink language syntax automatically. The spreadsheet also allows us to quantify the extracted sentences, compare them to the expected sentences and analyze if they are correct according to the process model. At last, in the third step (S3), we built a survey that mixes close and open-ended questions.

We \textbf{treated and analyzed data} using hybrid approaches (quantitative and qualitative). We used descriptive and inferential statistics for quantitative data, with support of the \textit{R Statistics 4.0.2} (for data analysis). We defined a 5\% significance level (\textbf{\textit{alpha=0.05}}) for the inferential statistics necessary to interpret the results. For qualitative data, we used the \textit{Atlas.ti 9} to analyze the participant's answers.

It is fundamental to determine how valid the research results are in scientific studies. It is needed to analyze all threats related to the building of the study. Creswell and Clark \citep{creswell2017designing} defined four main threats: conclusion, internal, construction, and external. Based on that, Table \ref{tab:threats} presents the main \textbf{threats of validity} in this research.

% \usepackage{multirow}
\begin{table}[ht!]
\centering
\caption{Threats of validity}
\label{tab:threats}
\small
%\resizebox{\textwidth}{!}{
\begin{tabular}{|c|p{4.0cm}|p{5.5cm}|}
\hline
\textbf{Type} &
  \textbf{Threat} &
  \textbf{Threat   Treatment} \\ \hline
\multirow{2}{*}{\textbf{Conclusion}} &
  Power   of statistical method analysis due to the existence of many statistical   methods and ways to use. &
  We   applied specific statistical methods according to each situation, such as   scales, type of variable, data behavior, etc. \\ \cline{2-3} 
 &
  Bias in the data selection due to the wishes of   the researchers. &
  We   published all data to allow the peer review and the study repeatability. \\ \hline
\multirow{4}{*}{\textbf{Internal}} &
  Lack   of participant training &
  According   to the study steps, we submitted all participants to training about using the   SYP method in detail. \\ \cline{2-3} 
 &
 Participants' activities historic &
  Although we presented the SYP method in the training step, it is only used for participants in the stage of execution of the study. \\ \cline{2-3} 
 &
  Participant's fatigue due to the study time &
  We   fixed the study in 90 minutes maximum. \\ \cline{2-3} 
 &
  Exchange information among participants &
  We applied the study in a single day, and we separated the participants. \\ \hline
\multirow{2}{*}{\textbf{Construction}} &
  Researcher's   expectations &
  The   researcher did not have contact with the participants in the study. \\ \cline{2-3} 
 &
  Improper instrumentation &
  We   submitted the metrics and questionnaire items to other researchers and   narratives specialists to guarantee the quality of the instrumentation. \\ \hline
\multirow{3}{*}{\textbf{External}} &
  The complexity of the process model. &
  Even though our process model was not complex, we used almost all BPMN symbols to simulate every possible extraction using the SYP method. However, we recognize the need to investigate using more complex and extensive models. We admit it is such as a research limitation, and we expect to look into it in future works. \\ \cline{2-3} 
 &
  Adequate   use of research methodologies. &
  To decrease it, we follow quasi-experimental methods available in the scientific literature because it is less controlled than experimentations, which is suitable for exploratory studies but keeps scientific rigor. \\ \cline{2-3} 
 &
  Result generalization &
  We   used screenwriters, the target audience, as participants. However, we know   that evaluating with more experienced screenwriters is necessary. \\ \hline
\end{tabular}
%}
\end{table}

\subsection{Study Execution}
\label{subsec:execucao_estudo}

In November 2022, we performed a quasi-experimental study with 19 participants. At the beginning of the study, all participants received the business process model in BPMN, a guide to using the SYP method and, a spreadsheet to execute the SYP method steps. The BPMN model contained \textbf{26 flow elements} (\textit{events, activities, and gateways}), and all BPMN elements create narratives from SYP’s extraction. Therefore, we expected the participants to extract all 26 sentences using the SYP method.

An instructor presented the SYP method to the participants, demonstrating its steps and how to use it (S1 – training). Next, the instructor asked participants to start using the SYP method from the BPMN and put the sentences into the spreadsheet (S2 - execution). After that, all participants should save the spreadsheets in a cloud space\footnote{Research data: \url{https://bit.ly/2UnnOEr}}. Finally, the participants answered a survey about their perceptions of the SYP method (S3 - survey). After that, we concluded the study.

\subsection{Data Analysis and Interpretation}
\label{subsec:interpretacao_estudo}

Our research investigated a group of 19 students between the ages of 17 and 35 who were enrolled in the foundations of information systems course as part of their bachelor's degree in information systems at a federal university in Brazil. It is important to mention that at that time, the course did not include teachings on business process modeling, making those students an acceptable target audience.

Table \ref{tab:analise_descritiva} summarizes participants' data and descriptive statistics (average, mode, standard deviation and percent) used to answer research questions. The table columns show the code of participants, the quantity and percent of extracted sentences, the quantity and percent of correct sentences, and the calculations of metrics MQ1 and MQ2.

\begin{table}[htbp]
\centering
\caption{Data summarizing - extracted and correct sentence metrics.}
\label{tab:analise_descritiva}
\small
%\resizebox{\textwidth}{!}{
\begin{tabular}{|c|c|c|c|c|}
\hline
\textbf{\textbf{Participant}} &
  \textbf{\begin{tabular}[c]{@{}c@{}}Extracted \\Sentences \\(QtdExt) \end{tabular}} &
  \textbf{\begin{tabular}[c]{@{}c@{}}Correct \\Sentences \\(QtdCorr)\end{tabular}} &
  \textbf{\begin{tabular}[c]{@{}c@{}}Metric MQ1 \\(QtdExt/QtdExp) \end{tabular}} &
  \textbf{\begin{tabular}[c]{@{}c@{}}Metric MQ2 \\(QtdCorr/QtdExp) \end{tabular}} \\ \hline
01             & 26 (100\%)            & 26 (100\%)            & 1.0 (100\%)           & 1.0 (100\%)          \\ \hline
03             & 26 (100\%)            & 26 (100\%)            & 1.0 (100\%)           & 1.0 (100\%)          \\ \hline
06             & 26 (100\%)            & 25 (96\%)             & 1.0 (100\%)           & 0.96 (100\%)         \\ \hline
08             & 26 (100\%)            & 26 (100\%)            & 1.0 (100\%)           & 1.0 (100\%)          \\ \hline
10             & 26 (100\%)            & 25 (100\%)            & 1.0 (100\%)           & 1.0 (100\%)          \\ \hline
11             & 26 (100\%)            & 26 (100\%)            & 1.0 (100\%)           & 1.0 (100\%)          \\ \hline
15             & 26 (100\%)            & 26 (100\%)            & 1.0 (100\%)           & 1.0 (100\%)          \\ \hline
17             & 26 (100\%)            & 25 (96\%)             & 1.0 (100\%)           & 0.96 (96\%)          \\ \hline
19             & 26 (100\%)            & 26 (100\%)            & 1.0 (100\%)           & 1.0 (100\%)          \\ \hline
02             & 26 (100\%)            & 26 (100\%)            & 1.0 (100\%)           & 1.0 (100\%)          \\ \hline
04             & 26 (100\%)            & 26 (100\%)            & 1.0 (100\%)           & 1.0 (100\%)          \\ \hline
05             & 26 (100\%)            & 26 (100\%)            & 1.0 (100\%)           & 1.0 (100\%)          \\ \hline
07             & 24 (92.31\%)          & 5 (21\%)              & 0.92 (92\%)           & 0.19 (19\%)          \\ \hline
09             & 26 (100\%)            & 26 (100\%)            & 1.0 (100\%)           & 1.0 (100\%)          \\ \hline
12             & 26 (100\%)            & 3 (12\%)              & 1.0 (100\%)           & 0.12 (12\%)          \\ \hline
13             & 26 (100\%)            & 6 (23\%)              & 1.0 (100\%)           & 0.23 (23\%)          \\ \hline
14             & 26 (100\%)            & 26 (100\%)            & 1.0 (100\%)           & 1.0 (100\%)         \\ \hline
16             & 26 (100\%)            & 26 (100\%)            & 1.0 (100\%)           & 1.0 (100\%)         \\ \hline
18             & 18 (69.23\%)          & 18 (100\%)            & 0.69 (69\%)           & 0.69 (100\%)         \\ \hline

\textbf{Average}  & \textbf{25.50 (98\%)} & \textbf{21.50 (84\%)} & \textbf{0.98 (98\%)}  & \textbf{0.86 (86\%)} \\ \hline
\textbf{Mode}  & \textbf{26 (100\%)} & \textbf{26 (100\%)} & \textbf{1.0 (100\%)}  & \textbf{1.0 (100\%)} \\ \hline
\textbf{\begin{tabular}[c]{@{}c@{}}Standard \\Deviation \end{tabular}} & \textbf{1.86 (7.1\%)} & \textbf{7.96 (30\%)} & \textbf{0.07 (7\%)}  & \textbf{0.30 (30\%)} \\ \hline
\end{tabular}
%}
\end{table}

Observing Table \ref{tab:analise_descritiva} is possible to perceive that most of the participants (mode column) could extract all of the sentences, and these sentences were correct related to the business process model. The average of extracted sentences was 25.5, corresponding to 98\% of them. The average number of correct sentences (21.5) was smaller than extracted sentences, corresponding to 84\% of them. We perceived that some participants (codes 5, 3, and 6), even extracting all sentences, most of them were not right. Metrics M1 and M2 presented, in general, values among 98\% and 86\%, respectively. These results show that the SYP method allows screenwriters to create complete and correct interactive narratives related to a business process model.

Using inferential statistics in the values of the metric MQ1, we analyzed and interpreted the hypothesis defined in Table \ref{tab:hypothesis}-Q1. We applied statistical tests to compare the measured values to a constant value \citep{nachar2008mann}. For the constant value, we use the value \textbf{1.0} once it is the maximum value for both metrics (MQ1 and MQ2) and, we expected 100\% of extracted and correct sentences. We want to highlight that we did not compare answers from two groups using different treatments (previous and post-tests) because we did not use other treatments. 

In studies that use statistical tests for inference is necessary to apply tests according to the scale of variables and the sample values. Thus, the first step is to verify whether sample data follow the data normality behavior. For MQ1, the adequate test is the \textit{Shapiro-Wilk} test because it is indicated for samples with few items \citep{shapiro1965analysis}. From the normality test, we observed that the values of MQ1 pointed to non-normality behavior (\textbf{\textit{p-value=3.97e-08} $<$ alpha=0.05}). Due to this result, the adequate test for evaluating the hypothesis is \textit{Wilcoxon's} test \citep{nachar2008mann}. Applying the test, we \textbf{accepted the alternative hypothesis}. It means that, with at least 95\% of certainty, \textbf{the SYP method CAN COMPLETELY extract elements from a BPMN model and put them into narrative sentences} (\textbf{\textit{p-value=5.30e-05 $<$ alpha=0.05}}).

Following the same analysis method for MQ2, we could evaluate its hypothesis (Table \ref{tab:hypothesis}-Q2). As MQ1, the \textit{Shapiro-Wilk's} test is adequate for testing the data normality behavior. As a result, we observed that the values of MQ1 do not follow the normality behavior (\textbf{\textit{p-value=1.18e-06} $<$ alpha=0.05}), which means that \textit{Wilcoxon's} test is adequate to check its hypothesis. \textit{Wilcoxon's} test resulted in the \textbf{acceptance of the alternative hypothesis} for MQ2. It means that, with 95\% of confidence, \textbf{Using the SYP method, screenwriters CAN CORRECTLY create interactive narratives from a business process model} (\textbf{\textit{p-value=7.76e-05 $<$ alpha=0.05}}).

\subsection{Answering Research Questions}
\label{subsec:desicussion}

We present some discussions on the research questions.

\begin{enumerate}[label={\bfseries Q\arabic*)}]
    \item \textbf{Could the method completely extract elements from a business process model and put them into narrative sentences?}
    
    In the quasi-experimental study, we evaluated the process of designing an interactive narrative from a business process model using the SYP method. We observed that screenwriters could extract all sentences from a business process model and create beat sheets from the data and statistical analysis.

    The qualitative analysis of participants' answers allows us to understand better their perceptions about it. In the survey, we asked:  ``What was your perception of the SYP's complexity to extract sentences? (easy, moderate, or complex). Why?''. Only one participant (5\%) indicated that the SYP is hard to run. He justifies it due to \textit{``[...] ... the number of actions and time-consuming filling the sentences in the spreadsheet''}. Seven participants (37\%) considered a moderate difficulty and, among the there was the perception of \textit{``[...] the attention to method details because if we make some mistake, the mistake follows to all other sentences''}. On the other hand, most participants (58\%) considered the SYP method easy to extract sentences.
    
    Therefore, the quantitative and quantitative analysis reach similar evidence: the SYP method was considered a method that allows extracting all sentences from a business process model. These results go straight to the SYP method to provide a direct way for screenwriters to design business process model-based interactive narratives without a solid knowledge of the BPMN language.\newline
    
    \item \textbf{Does the method support screenwriters creating correct interactive narratives according to a business process model?}
    
    Business processes model-based interactive narrative must align narrative elements to process model elements. In other words, it needs to be reliable related to the business process. Hence, screenwriters must guarantee the trust of the narrative by extracting elements from the process model, creating sentences, and putting them into a coherent sequence according to the model. In our quasi-experimental study, the statistical analysis showed that the SYP method supported screenwriters to perform this task.

    We also looked to qualitative questions to complement the quantitative analyses. Thus, the survey presented the question: ``Did the SYP method help create a step outline representing the business process model? (yes, somewhat, no). Why?'' That question received 100\% of affirmative answers (yes). To illustrate that, we brought a participant's quotation: \textit{``I believe that building narrative scripts from the BPMN model guarantee fidelity of the narrative to the process model. Developing it without support is a risk because it is hard to say that the script will follow the process model. This gap between model and narratives decreases when we use the method (SYP)''}.

    The SYP can support screenwriters to extract correctly sentences aligned to business process models. Qualitative and quantitative results evidence it. Hence, we can answer the questions affirmatively.
\end{enumerate}    

Finally, we can analyze the research problem: \textbf{\textit{How to systematize the business process-based interactive narratives design?}} We observed evidence that the SYP method is a possible approach for that. The SYP is composed of systematic steps that help screenwriters turn business process models into narratives. 

Our research contributes to business process management (including social BPM) and interactive narrative design. We provide an alternative way to model and present business processes that help people understand them simpler and contribute to organizational improvements. We also provide a systematic method to organize sentences in step outlines and put them into an interactive narrative. 

\subsubsection{Limitations}\leavevmode

\noindent Concerning limitations, our evaluation was a controlled study. We used participants with some experience to create narratives but, they were not professionals. Thus, we can say that results seem valid in the context of this study, but we believe there is a need to evaluate with more experienced screenwriters in the future.

The BPMN model used in the study is another study limitation. In evaluation, we used a small model with 26 elements. Although we observed evidence of the viability of using the SYP method in this study, we believe in the need to evaluate the SYP with more extensive and more complex models.

Lastly, in this study, the results of the SYP method were the beat sheets. With the beat sheets made by the SYP method, we could run as an interactive narrative in the Inkle. However, the study participants did not approach all narrative elements (complexity of characters, places, narrator, plots, acts, time, and others). Thus, we think that still are a need to move forward in this research and, in fact, create full interactive narratives.

In DSR-based research, the scientific investigations are cyclic. The limitations and challenges found in evaluations will serve as future \textit{insights} in new investigative cycles. It is why it is so important to present those research limitations.

\section{Conclusion}
\label{sec:conclusion}

Some research proposals on business processes modeling use narrative techniques to discover and identify process elements from the textual content. This article looks at that approach from another perspective: creating interactive narratives from business processes models. We understood that interactive narratives could present and represent business process models simply and attractively for ordinary people, not only processes managers.

Usually, the processes modeling languages give symbols for easy understanding to people in the organizational management. This presumed ease can not be valid if we think of ordinary people who do not know BPM. There are screenwriters in the middle of those ordinary people who will design narratives from business processes models.

In this work, we presented and evaluated the SYP method. The SYP is a systematic proposal to support screenwriters in designing interactive narratives based on business process models. Screenwriters do not necessarily need to know the meaning of model elements to create interactive narratives using the SYP. The result of the SYP method is interactive narratives that represent the business process model with fidelity.

An alternative approach based on hypermedia content can help people understand and improve organizational processes through collaboration and immersion. We evaluated the SYP method using a quasi-experimental study, analyzing if screenwriters could extract completely and correctly sentences from a business process model using it. As a result, we observed affirmative evidence that the SYP method enabled it. Therefore, even approaching narratives in the form of beat sheets, we can say there is evidence that the SYP method is a possible proposal to reach our research problem.

This article presented the first DSR cycle of more extensive research. Hence, the limitation and challenges found in the artifact evaluation are new requirements to other scientific cycles and, consequently, future works. Thus, the next DSR cycle will approach full interactive narratives, considering details of characters, acts, plots, and screenwriters' creativity to explore ludic aspects of business processes. Additionally, we want to develop software to automatize the manual steps of the SYP. Finally, we would like to evaluate the method with professionals and more complex business processes models.

%\section*{Acknowledgement}
%The project is funded in part by the National Institutes of Health, under Grant No. 5R01CA136535.

%%%%%%%%%%%%%%%%%%%%%%%%%%%%%%%%%%%%%%%%%%%%%%%%%%%%%%%%%%%%%%%%%%%%%%%%%%%%%%%%%%%%%

\bibliographystyle{agsm}
\bibliography{References/references}

@String{Computing = "Computing" }

@String{Computer = "{IEEE} Computer" }

@String{Springer = "Springer-Verlag" }

@techreport{basili1992software,
  title={Software modeling and measurement: the Goal/Question/Metric paradigm},
  author={Basili, Victor R},
  year={1992},
  institution={University of Maryland},
  number={CS-TR-2956, UMIACS-TR-92-9}
}

@book{dumas2018fundamentals,
  title={Fundamentals of business process management},
  author={Dumas, Marlon and La Rosa, Marcello and Mendling, Jan and Reijers, Hajo A and others},
  volume={2},
  year={2018},
  publisher={Springer},
  address={Heidelberg}
}

@book{object_management_group_inc_omg_business_2014,
	edition = {2.0.2},
	title = {Business Process Model and Notation ({BPMN})},
	publisher = {Object Management Group ({OMG})},
	author = "{Object Management Group}",
	year = {2014},
	address={Massachusetts}
}

@inproceedings{aarseth_narrative_2012,
	location = {North Carolina},
	title = {A Narrative Theory of Games},
	pages = {129--133},
	booktitle = {International Conference on the Foundations of Digital Games},
	author = {Aarseth, E.},
	publisher = {Association for Computing Machinery},
	address = {New York, NY, USA},
	year = {2012},
}

@inbook{ryan2017narrative,
    author = {Ryan, Marie-Laure},
    publisher = {John Wiley \& Sons, Ltd},
    isbn = {9781118472262},
    title = {Narrative},
    booktitle = {A Companion to Critical and Cultural Theory},
    chapter = {33},
    pages = {517-530},
    year = {2017},
    address={New Jersey}
}

@article{riedl2013interactive,
  title={Interactive narrative: An intelligent systems approach},
  author={Riedl, Mark Owen and Bulitko, Vadim},
  journal={Ai Magazine},
  volume={34},
  number={1},
  pages={67--67},
  year={2013}
}

@inproceedings{goncalves_collaborative_2010,
	location = {Funchal, Madeira, Portugal},
	author = {Gonçalves, João Carlos A. R. and Santoro, Flávia and Baião, Fernanda},
	title = {Collaborative Business Process Elicitation through Group Storytelling},
	isbn = {978-989-8425-04-1 978-989-8425-05-8 978-989-8425-06-5 978-989-8425-07-2 978-989-8425-08-9},	
	doi = {10.5220/0002910002950300},
	eventtitle = {12th International Conference on Enterprise Information Systems},
	pages = {295--300},
	booktitle = {Proceedings of the 12th International Conference on Enterprise Information Systems},
	publisher = {{SciTePress} - Science and and Technology Publications},
	urldate = {2021-07-12},
	year = {2010a},
	address={Portugal}
}

@inproceedings{de_a_r_goncalves_case_2010,
	location = {Shanghai, China},
	title = {A case study on designing business processes based on collaborative and mining approaches},
	isbn = {978-1-4244-6763-1},	
	doi = {10.1109/CSCWD.2010.5471899},
	eventtitle = {2010 14th International Conference on Computer Supported Cooperative Work in Design ({CSCWD})},
	pages = {611--616},
	booktitle = {The 2010 14th International Conference on Computer Supported Cooperative Work in Design},
	publisher = {{IEEE}},
	author = {Gonçalves, Joao Carlos de A. R. and Santoro, Flavia Maria and Baiao, Fernanda Araujo},
	urldate = {2021-07-12},
	year = {2010b},
	address={China}
}

@article{santoro_acquiring_2010,
	title = {Acquiring knowledge on business processes from stakeholders’ stories},
	volume = {24},
	issn = {14740346},
	doi = {10.1016/j.aei.2009.07.002},
	pages = {138--148},
	number = {2},
	journal = {Advanced Engineering Informatics},
	author = {Santoro, Flávia Maria and Borges, Marcos R.S. and Pino, José A.},
	urldate = {2021-07-12},
	year = {2010},
	langid = {english},
}

@book{campbell2015experimental,
  title={Experimental and quasi-experimental designs for research},
  author={Campbell, Donald T and Stanley, Julian C},
  year={2015},
  publisher={Ravenio Books},
  address={Stanford}
}

@inproceedings{juchova_storyboards_2010,
	title = {Storyboards in business process modeling},
	booktitle = {International Industrial Simulation Conference},
	author = {Juchova, Veronika and Stolfa, Svatopluk and Ježek, David and Vondrak, Ivo},
	year = {2010},
	publisher={EUROSIS},
	address={Ghent},
	pages={57--61}
}

@inproceedings{donikian2004writing,
  title={Writing interactive fiction scenarii with DraMachina},
  author={Donikian, St{\'e}phane and Portugal, Jean-No{\"e}l},
  booktitle={International conference on technologies for interactive digital storytelling and entertainment},
  pages={101--112},
  year={2004},
  publisher = {Springer},
  address = {Berlin}
}

@book{montfort2005twisty,
  title={Twisty Little Passages: an approach to interactive fiction},
  author={Montfort, Nick},
  year={2005},
  publisher={Mit Press},
  address={London}
}

@article{luo2015review,
  title={A review of interactive narrative systems and technologies: a training perspective},
  author={Luo, Linbo and Cai, Wentong and Zhou, Suiping and Lees, Michael and Yin, Haiyan},
  journal={Simulation},
  volume={91},
  number={2},
  pages={126--147},
  year={2015},
  publisher={Sage Publications Sage UK: London, England}
}

@book{malkewitz2006technologies,
  title={Technologies for interactive digital storytelling and entertainment},
  author={Malkewitz, Stefan G{\"o}bel Rainer and Iurgel, Ido},
  year={2006},
  publisher={Springer},
  address={Germany}
}

@article{de2011let,
  title={Let Me Tell You a Story-On How to Build Process Models},
  author={AR Gon{\c{c}}alves, Jo{\~a}o Carlos and Santoro, Fl{\'a}via Maria and Bai{\~a}o, Fernanda Araujo},
  journal={J. Univers. Comput. Sci.},
  volume={17},
  number={2},
  pages={276--295},
  year={2011}
}

@article{antunes2020eliciting,
  title={Eliciting process knowledge through process stories},
  author={Antunes, Pedro and Pino, Jose A and Tate, Mary and Barros, Alistair},
  journal={Information Systems Frontiers},
  volume={22},
  number={5},
  pages={1179--1201},
  year={2020},
  publisher={Springer}
}

@article{hevner2004design,
  title={Design science in information systems research},
  author={Hevner, Alan R and March, Salvatore T and Park, Jinsoo and Ram, Sudha},
  journal={MIS quarterly},
  pages={75--105},
  year={2004},
  publisher={JSTOR},
  volume=28,
  number=1
}

@article{pimentel2020design,
  title={Design Science Research: scientific research connected to artifacts design},
  author={Pimentel, Mariano and Filippo, Denise and Santos, Thiago Marcondes},
  journal={RE@D--Journal of Distance Education and eLearning},
  volume={3},
  number={1},
  pages={37--61},
  year={2020}
}

@article{kim2016quasi,
  title={Quasi-experimental designs for causal inference},
  author={Kim, Yongnam and Steiner, Peter},
  journal={Educational psychologist},
  volume={51},
  number={3-4},
  pages={395--405},
  year={2016},
  publisher={Taylor \& Francis}
}

@book{creswell2017designing,
  title={Designing and conducting mixed methods research},
  author={Creswell, John W and Clark, Vicki L Plano},
  year={2017},
  publisher={Sage publications},
  address={Los Angeles}
}

@article{nachar2008mann,
  title={The Mann-Whitney U: A test for assessing whether two independent samples come from the same distribution},
  author={Nachar, Nadim and others},
  journal={Tutorials in quantitative Methods for Psychology},
  volume={4},
  number={1},
  pages={13--20},
  year={2008}
}

@article{shapiro1965analysis,
  title={An analysis of variance test for normality (complete samples)},
  author={Shapiro, Samuel Sanford and Wilk, Martin B},
  journal={Biometrika},
  volume={52},
  number={3/4},
  pages={591--611},
  year={1965},
  publisher={JSTOR}
}

@book{montfort2007generating,
  title={Generating narrative variation in interactive fiction},
  author={Montfort, Nick},
  year={2007},
  publisher={University of Pennsylvania},
  address={Pennsylvania}
}

@inproceedings{green2019novella,
  title={Novella 2.0: a hypertextual architecture for interactive narrative in games},
  author={Green, Daniel and Hargood, Charlie and Charles, Fred},
  booktitle={Proceedings of the 30th ACM Conference on Hypertext and Social Media},
  pages={77--86},
  year={2019},
  publisher = {Association for Computing Machinery},
  address = {New York, NY, USA}
}

@article{rodriguez2014science,
  title={Science on television: audiences, markets and authority. Some conclusions},
  author={Rodr{\'\i}guez, Clara Florensa and Hochadel, Oliver and Tabernero, Carlos},
  journal={Actes d'Hist{\`o}ria de la Ci{\`e}ncia i de la T{\`e}cnica},
  pages={127--136},
  year={2014},
  volume={7},
  number={1}
}

@article{classeEtal2021,
author = {Classe, Tadeu Moreira and De Araujo, Renata Mendes and Xex\'{e}o, Geraldo Bonorino and Siqueira, Sean Wolfgand Matsui},
title = {Public Processes Are Open for Play},
year = {2021},
issue_date = {October 2021},
publisher = {Association for Computing Machinery},
address = {New York, NY, USA},
volume = {2},
number = {4},
issn = {2691-199X},
doi = {10.1145/3474879},
journal = {Digit. Gov.: Res. Pract.},
month = {dec},
articleno = {32},
numpages = {18}
}

@article{engiel2014designing,
  title={Designing public service process models for understandability},
  author={Engiel, Priscila and Araujo, Renata and Cappelli, Claudia},
  journal={Electronic Journal of e-Government},
  volume={12},
  number={1},
  pages={pp95--111},
  year={2014}
}

@inproceedings{giacomo2015declarative,
  title={Declarative process modeling in BPMN},
  author={Giacomo, Giuseppe De and Dumas, Marlon and Maggi, Fabrizio Maria and Montali, Marco},
  booktitle={International conference on advanced information systems engineering},
  pages={84--100},
  year={2015},
  organization={Springer}
}

@book{mcerlean2018interactive,
  title={Interactive narratives and transmedia storytelling: Creating immersive stories across new media platforms},
  author={McErlean, Kelly},
  year={2018},
  publisher={Routledge}
}

@inproceedings{piredda2015social,
  title={Social media fiction},
  author={Piredda, Francesca and Ciancia, Mariana and Venditti, Simona},
  booktitle={International Conference on Interactive Digital Storytelling},
  pages={309--320},
  year={2015},
  organization={Springer}
}

@inproceedings{brambilla2012combining,
  title={Combining social web and BPM for improving enterprise performances: the BPM4People approach to social BPM},
  author={Brambilla, Marco and Fraternali, Piero and Vaca Ruiz, Carmen Karina},
  booktitle={Proceedings of the 21st International Conference on World Wide Web},
  pages={223--226},
  year={2012}
}

\end{document}